# A Class of Practical and Acceptable Social Welfare Orderings That Satisfy the Principles of Aggregation and Non-Aggregation: Reexamination of the Tyrannies of Aggregation and Non-Aggregation

Norihito Sakamoto*

## Abstract

This paper revisits impossibility results on the tyrannies of aggregation and non-aggregation. I propose two aggregation principles --quantitative aggregation and ratio aggregation—and investigate theoretical implications. As a result, I show that

---

* Tokyo University of Science, E-mail: n-sakamoto@rs.tus.ac.jp

This study is a part of joint international research projects with CPNSS at LSE.; I sincerely thank Alex Voorhoeve for his friendship and arrangements for our joint research projects. I would like to thank Geir Asheim, Marc Fleurbaey, Iwao Hirose, John Weymark, Yongsheng Xu, and Stephane Zuber for their kind comments and suggestions, which were very helpful in revising the paper. I also offer my special thanks for the grants from Fostering Joint International Research (B) (JSPS KAKENHI Grant Number: 20KK0036).

quantitative aggregation and minimal non-aggregation are incompatible while ratio aggregation and minimal non-aggregation are compatible under the assumption of standard axioms in social choice theory. Furthermore, this study provides a new characterization of the leximin rule by using replication invariance and the strong version of non-aggregation. Finally, I propose a class of practical and acceptable social welfare orderings that satisfy the principles of aggregation and non-aggregation, which has various advantages over the standard rank-discounted generalized utilitarianism.

# 1. Introduction

This study reexamines the compatibility problem of the principles of aggregation and non-aggregation. The *aggregation principle* requires that a tiny loss of one person should not have priority over great gains of all the rest. This is a sound and important property in any social welfare decision. It is hardly for any rational social planner to force all the rest to suffer from great sacrifices for the sake of one individual's tiny gain. Furthermore, a society that makes majority's great sacrifices for the sake of only one individual's small gain cannot be stable, as the majority can even ostracize that one person.[1]

On the other hand, the *non-aggregation principle* requires that a great gain of one poor individual should have priority over tiny losses of many individuals. This condition is also an important property for a sound and liberal society. No individual should make a great sacrifice for the sake of very tiny gains for majorities.[2] As a practical matter, the requirement of non-aggregation is of great significance in various contexts such as the protection of the rights of ethnic minorities, reasonable accommodation for persons with disabilities, considerations of patients with incurable and rare diseases which are ignored in usual cost-effectiveness analysis, and mediation between the interests of the current and future generations.[3] For example, a very expensive therapeutic drug greatly improves quality of lives of a small number of patients with intractable diseases, it should be said to be socially good that a society shares the cost of the drug and everyone has a small loss.[4]

---

[1] Imposing an *excess burden* on people around a person with disabilities for the sake of a small gain is generally not considered a *reasonable accommodation*. Furthermore, it is difficult to accept a situation in which everyone incurs huge losses in order to improve the living standards of the poor by a tiny amount.

[2] See Scanlon (1998, p. 235): Suppose that Jones has suffered an accident in the transmitter room of a television station. Electrical equipment has fallen on his arm, and we cannot rescue him without turning off the transmitter for fifteen minutes. A World Cup match is in progress, watched by many people, and it will not be over for an hour. Jones's injury will not get any worse if we wait, but his hand has been mashed and he is receiving extremely painful electrical shocks. Should we rescue him now or wait until the match is over? Does the right thing to do depend on how many people are watching—whether it is one million or five million or a hundred million? It seems to me that we should not wait, no matter how many viewers there are...

However, it is known that these very moderate principles are generally not well reconciled. In their celebrated paper, Fleurbaey and Tungodden (2010) showed that there is no social quasi-ordering that satisfies the aggregation principle, the non-aggregation principle, replication invariance, weak Pareto, and Pigou-Dalton transfer. Their results suggest that in order for the principles of aggregation and non-aggregation to be compatible, replication invariance must be dropped. In fact, they propose the *geometric*

---

[3] Note that interests of the current generation are often too much estimated in the simple discounted utilitarian approach. Recently, Zuber and Asheim (2012) and Asheim and Zuber (2014) propose and apply the rank-discounted social welfare function to the context of intergenerational evaluation or variable populations, but as shown in this paper, this social welfare function has the flaw of protecting the interests of minorities too much and neglecting the interests of the whole. In fact, balancing of interests between majorities and minorities seems to be well supported by various economic experiments. See Cowell et all. (2015), Schumacher et al. (2017), Voorhoeve et al. (2019), and Luptakova and Voorhoeve (2023).

[4] In the context of classical economics, the non-aggregation principle also concerns the protection of domestic industry, which is undermined by the promotion of free trade. Does the non-aggregation principle require protecting domestic industry due to the fact that promoting free trade comes at great cost to workers in domestic industry? However, if it is potentially Pareto-superior to promote free trade than to protect domestic industries, then by redistributing the gains of trade and encouraging domestic workers to move to growing industries, a policy maker can solve a conflict between the principles of Pareto and non-aggregation.

*Gini ordering* as an example of a possibility result.[5] They further showed that the combination of non-aggregation, replication invariance, weak Pareto, and Pigou-Dalton transfer must yield a refinement of the maximin rule.

This study supplements this previous study with some impossibility/ possibility results and proposes a class of new practical and acceptable social orderings that satisfies both the principles of aggregation and non-aggregation. Specifically, this study proposes two new principles of aggregation. The first one, the *quantitative aggregation principle*, requires that a small loss of one individual should not have priority over sufficiently large gains of a specific number of *m* or more individuals. It will then be shown that this aggregation principle is incompatible with the non-aggregation principle, even without imposing replication invariance and Pigou-Dalton transfer. The second one, the *ratio aggregation principle*, requires that a small loss of one individual should not have priority over sufficiently large gains of a sufficiently large proportion of the total population. This aggregation principle will be shown to be compatible with the non-aggregation principle, strong Pareto, and anonymity (although requiring replication invariance yields an impossibility theorem). It will also be shown that a social ordering proposed in this paper has features superior to the geometric Gini ordering, which is a compromised solution between the traditional non-aggregation and aggregation principles.

Furthermore, this study shows that a social quasi-ordering that satisfies anonymity, strong Pareto, replication invariance, and a strong version of the non-aggregation principle must be leximin. Thus, requiring both the strong non-aggregation principle and replication invariance leads to give absolute priority to the relatively poor. In other words, the combination of the strong non-aggregation principle and replication invariance unintentionally yields a call for extreme fairness, such as Hammond equity.

---

[5] In addition, *nontransitive* social evaluation methods that respect the non-aggregation principle have also been proposed. For details, see Scanlon (1998, Ch. 5), Fleurbaey et al. (2009), and Voorhoeve (2014).

Main contributions in this paper are as follows. First, this study proposes two aggregation principles and shows that the principles of quantitative aggregation and non-aggregation are incompatible, even without imposing replication invariance. Second, I show that the principles of ratio aggregation and non-aggregation are compatible with standard axioms. Third, this study obtains a new characterization of the leximin rule in which the only social quasi-ordering satisfying anonymity, strong Pareto, replication invariance and the strong non-aggregation principle is leximin. Fourth, I examine the problems of rank-discounted generalized utilitarianism (a generalization of the geometric Gini ordering) and propose a class of practical and acceptable social orderings that satisfies both the principles of aggregation and non-aggregation.

# 2. Notation and Definitions

This section explains basic notation, definitions, and axioms. The set of individuals $N = \{1, \ldots, n\}$ is assumed to be variable and finite ($n > 2$). Suppose that well-being of individuals can be represented by real values and be cardinally interpersonal comparable. Given a set of individuals $N = \{1, \ldots, n\}$, a profile of individual well-being $u_N = (u_1, \ldots, u_n)$ is represented by a vector of *n*-tuple real values. Also, for any well-being profile $u_N$, let $u_{[N]} = (u_{[1]}, \ldots, u_{[n]})$ be a rearranged profile of individual well-being in decreasing order, that is, $u_{[1]} \leq \cdots \leq u_{[n]}$. For any natural number *k* and any profile $u_N$, let $k * u_N$ be the *k* replications of profile $u_N$. I write $U = \bigcup_{n>2} \mathbb{R}^n$ for the universal set of profiles of individual well-being with variable population size. The aim of this paper is to obtain a class of acceptable and practical social orderings $\succcurlyeq$ on the set of profiles that satisfy desirable and standard axioms in social choice theory. A binary relation $\succcurlyeq$ defined on $U$ is a *social quasi-ordering* if it satisfies reflexivity and transitivity. If a social quasi-ordering satisfies completeness, it is called a *social ordering*.[6] In general,

---

[6] Reflexibility, completeness, and transitivity of a binary relation $\succcurlyeq$ are defined as follows, respectively.

**Reflexibility.** $\forall u_N \in U, u_N \succcurlyeq u_N$.
**Completeness.** $\forall u_N, v_N \in U, u_N \succcurlyeq v_N$ or $v_N \succcurlyeq u_N$.
**Transitivity.** $\forall u_N, v_N, w_N \in U, u_N \succcurlyeq v_N$ and $v_N \succcurlyeq w_N \rightarrow u_N \succcurlyeq w_N$.

Note that these definitions are constrained in the restricted domain with the same population size (it is easily extended to the universal domain with variable populations but this extension does not be needed to obtain all main results in this paper).

$u_N \succcurlyeq v_N$ should be interpreted that $u_N$ is at least as socially good as $v_N$. Then, I use the normal notation as follows: $u_N \succ v_N \leftrightarrow u_N \succcurlyeq v_N$ & not $v_N \succcurlyeq u_N$ and $u_N \sim v_N \leftrightarrow u_N \succcurlyeq v_N$ & $v_N \succcurlyeq u_N$.

Next, let us consider standard axioms that a social ordering should satisfy. One of the standard axioms is *anonymity* which requires that well-being of individuals should be treated equally.

**Anonymity.**

$\forall u_N, v_N \in U, \forall \text{bijection } \pi \text{ on } N, \text{if } \forall i \in N, u_i = v_{\pi(i)}, \text{then } u_N \sim v_N$.

Let us introduce two efficiency requirements: the strong Pareto and weak Pareto principles. The strong Pareto axiom requires that making someone strictly better off without anyone being worse off should be always good for society. Also, it requires that a social ranking must respect Pareto indifference relations. The weak Pareto axiom requires that improving everyone's well-being should be always good for society. It is obvious that strong Pareto implies weak Pareto by definition.

**Strong Pareto.**

$\forall u_N, v_N \in U, \text{if } \forall i \in N, u_i \geq v_i, \text{then } u_N \succcurlyeq v_N$. Moreover, if $\forall i \in N, u_i \geq v_i$ and $\exists j \in N, u_j > v_j, \text{then } u_N \succ v_N$.

**Weak Pareto.**

$\forall u_N, v_N \in U, \text{if } \forall i \in N, u_i > v_i, \text{then } u_N \succ v_N$.

Let us define the Pigou-Dalton transfer axiom as a requirement of distributive justice. This axiom requires that reducing a gap in well-being between two people without the total sum changing should not decrease social welfare as long as the transfer keeps the well-being of others constant.

**Pigou-Dalton Transfer.**

$\forall u_N, v_N \in U, \forall \varepsilon \in \mathbb{N}$, if $\exists i, j \in N, u_i - \varepsilon = v_i \geq v_j = u_j + \varepsilon$ & $\forall h \in N \setminus \{i, j\}, u_h = v_h$, then $v_N \succcurlyeq u_N$.

Next, consider the replication invariance axiom as a consistency condition of social choice problems with variable population sizes. This axiom requires that social judgments at the original population size should be the same one in a *k*-replication economy. Note that this condition requires consistency of ranking on profiles with *the same population size*.

**Replication Invariance.**

$\forall u_N, v_N \in U, \forall k \in \mathbb{N}, \ u_N \succcurlyeq v_N \leftrightarrow k * u_N \succcurlyeq k * v_N$

Finally, I introduce the principles of aggregation and non-aggregation, which are main axioms of this study. The non-aggregation principle basically requires that one poor person should not be forced to make a great sacrifice for very small benefits of many rich persons. The following non-aggregation principle was introduced by Fleurbaey and Tungodden (2010), which requires that the small sacrifice $\beta$ of the richest individual with well-being above a threshold $\theta_r$ should be acceptable if the poorest individual with well-being below a threshold $\theta_p$ can obtain a sufficiently large benefit $\alpha$.

**Minimal Non-Aggregation (Fleurbaey and Tungodden 2010).**

$\exists \theta_r > \theta_p > 0$, $\exists \alpha > \beta > 0$, $\forall u_N, v_N \in U$ with $|N| = n$, $\forall i \in N, \forall M \subseteq N \setminus \{i\}$, if $[\theta_p \geq v_i = v_{[1]} \geq u_i + \alpha]$ & $[\forall j \in M, u_j = u_{[n]} \geq \theta_r$ & $v_j = v_{[n]} \geq u_j - \beta]$ & $[\forall h \in N \setminus (M \cup \{i\}), v_h = u_h]$, then $v_N \succcurlyeq u_N$.

Note that this condition says nothing about improvements in well-being of individuals who are not the worst-off but below $\theta_p$. This requirement is just considering the situation where there is one poorest person and all others are sufficiently rich or unaffected. This situation cannot reflect the non-aggregation problem proposed by Scanlon (1998) because potential benefits of persons below the threshold should be respected even if the person is not exactly the worst off among the people. Also, if the threshold of poverty $\theta_p$ is sufficiently high, the requirement of non-aggregation should be abstained or less important for aggregating losses and gains of individuals. Indeed, this study tries to examine just one implication of some variants of non-aggregation axioms without any thresholds. The following *strong non-aggregation* principle requires that no individual should be forced to make a great sacrifice for the slight benefits of many persons. Without considering any thresholds, combing this version of the non-aggregation principle with standard axioms (anonymity, strong Pareto, and replication invariance) leads to the leximin rule. Thus, it turns out that the strong non-aggregation principle is fundamentally incompatible with the aggregation principle.

**Strong Non-Aggregation.**

$\forall \alpha > 0, \exists \beta > 0$ with $\alpha > \beta, \forall u_N, v_N \in U, \forall i \in N, \forall M \subseteq N \setminus \{i\}$, if $\left[\forall j \in M, u_j - \beta = v_j > v_i = u_i + \alpha\right]$ & $[\forall h \in N \setminus (M \cup \{i\}), v_h = u_h]$, then $v_N \succcurlyeq u_N$.

Let us then introduce the aggregation principle.[7] The aggregation principle basically requires that the small sacrifice of a single person should be tolerable if a sufficiently large number of people can get great benefits enough. In this paper, I divide the basic requirement of this principle into two parts. The first one is the quantitative aggregation principle, which requires that the greater benefit $\gamma$ of well-being of *m* or more persons have priority over the small sacrifice $\delta$ of a single person. The second one is the ratio aggregation principle, which requires that the greater benefit $\gamma$ of well-being of $100\lambda\%$ or more of the population have priority over the small sacrifice $\delta$ of a single

---

[7] Fleurbaey and Tungodden (2010) consider the following minimal aggregation principle.

**Minimal Aggregation.**

$\exists n \in \mathbb{N}, \forall u_N, v_N \in U$ with $|N| = n, \exists \gamma > \delta > 0, \forall i \in N,$ if $[v_i \geq u_i - \delta]$ & $\left[\forall j \in N \setminus \{i\}, v_j \geq u_j + \gamma\right]$, then $v_N \succcurlyeq u_N$.

This condition seems too weak to aggregate gains and losses of well-being for all population sizes. It just requires that the aggregation matters for at least one population size *n*. Also, $\gamma$ and $\delta$ rely on the profiles, which allows the situation where the very small sacrifice of one individual overwhelms very large benefits of huge populations. As Fleurbaey and Tungodden (2010) pointed out, the author strongly agree that this should be substantially strengthened to the extent in which aggregation always matters for all population sizes and all situations.

person.

**Quantitative Aggregation.**

$\exists m \in \mathbb{N}$ with $m > 2$, $\exists \gamma > \delta > 0$, $\forall u_N, v_N \in U$ with $|N| = n > m$, $\forall i \in N$, if $[v_i \geq u_i - \delta]$ & $[\forall j \in M \subseteq N$ with $|M| \geq m$, $v_j \geq u_j + \gamma]$ & $[\forall h \in N \setminus (M \cup \{i\}), v_h = u_h]$, then $v_N \succcurlyeq u_N$.

**Ratio Aggregation.**

$\exists \lambda \in (0, 1)$, $\exists \gamma > \delta > 0$, $\forall u_N, v_N \in U$ with $|N| = n$, $\forall i \in N$, $\forall M \subseteq N$ with $|M| \geq [\lambda n]$, if $[v_i \geq u_i - \delta]$ & $[\forall j \in M, v_j \geq u_j + \gamma]$ & $[\forall h \in N \setminus (M \cup \{i\}), v_h = u_h]$, then $v_N \succcurlyeq u_N$, where $[\lambda n] = \min\{m \in \mathbb{N} | m \geq \lambda n\}$.

Note that the repeated *k*-time applications of the quantitative or ratio aggregation principle would make it socially acceptable to impose a severe sacrifice $k\gamma$ on a single person for the sake of an enormous benefit $k\delta$ of more than *m* individuals or more than $100\lambda$% of the population size. In this sense, both the principles of quantitative and ratio aggregation should be constrained on their ranges of well-being profiles to which they apply. For example, the aggregation principles should be applied for only profiles in a constrained domain $(\underline{u}, \overline{u})^n$. However, since this does not significantly change the possibility/impossibility results obtained in this study and only makes the argument somewhat more complicated, I will continue to use the simplified versions of the aggregation principles. Needless to say, the question of whether it is acceptable to force a single person to make a great sacrifice for the greater benefits of the majority is one of the vexing dilemmas of consequentialism.[8]

---

[8] One of famous examples is the survival lottery (Harrisson 1975). Is it good to kill a healthy man and distribute his organs to save the lives of five people in need of organ transplants? The trolley problem has a similar dilemma. In this sense, there is a clear problem with consequentialism and simple utilitarianism.

# 3. Impossibility and Possibility Results

In this section, I will prove some impossibility and possibility results concerning the principles of aggregation and non-aggregation. First, I will show that the principles of quantitative aggregation and non-aggregation are incompatible, even without imposing replication invariance.

**Proposition 1.** *There exists no social quasi-ordering that satisfies weak Pareto, quantitative aggregation, and minimal non-aggregation.*

Proof. Let $\overline{u} > \theta_r > \theta_p > \underline{u}$ and $N = \{1, \ldots, n\}$, where $\theta_p \geq \underline{u} + \alpha$. Choose $\overline{u}$ and natural numbers $h, l \in \mathbb{N}$ such that $n = hlm + l$, $\overline{u} - l\beta > \theta_r$, $\underline{u} > \underline{u} + \alpha - h\delta$, and $\overline{u} > \overline{u} + \gamma - l\beta$. Then, consider the following profiles.

$u_N = (l * \underline{u}, hlm * \overline{u})$;

$u_N^1 = (\underline{u} + \alpha, (l-1) * \underline{u}, hlm * (\overline{u} - \beta))$;

$u_N^2 = (\underline{u} + \alpha, \underline{u} + \alpha, (l-2) * \underline{u}, hlm * (\overline{u} - 2\beta))$;

$$\vdots$$

$u_N^l = (l * (\underline{u} + \alpha), hlm * (\overline{u} - l\beta))$;

$w_N^1 = (\underline{u} + \alpha - h\delta, (l-1) * (\underline{u} + \alpha), hm * (\overline{u} + \gamma - l\beta), hm(l-1) * (\overline{u} - l\beta))$;

$w_N^2 = (2 * (\underline{u} + \alpha - h\delta), (l-2) * (\underline{u} + \alpha), 2hm * (\overline{u} + \gamma - l\beta), hm(l-2) * (\overline{u} - l\beta))$;

$$\vdots$$

$$w_N^l = (l * (\underline{u} + \alpha - h\delta), hlm * (\overline{u} + \gamma - l\beta)).$$

By *l* applications of minimal non-aggregation, $u_N^l \succcurlyeq \cdots \succcurlyeq u_N^2 \succcurlyeq u_N^1 \succcurlyeq u_N$.

By *h* applications of quantitative aggregation, $w_N^1 \succcurlyeq u_N^l$.

By *l* applications of the above processes, $w_N^l \succcurlyeq \cdots \succcurlyeq w_N^2 \succcurlyeq w_N^1$.

By transitivity, $w_N^l \succcurlyeq u_N$.

However, weak Pareto implies $u_N \succ w_N^l$, which is a contradiction. ∎

Note that Proposition 1 does not require both Pigou-Dalton transfer and replication invariance dislike the impossibility result in Fleurbaey and Tungodden (2010). Generally, a combination of their minimal aggregation principle and replication invariance implies quantitative aggregation, but not vice versa. In this sense, Proposition 1 suggests that replication invariance and Pigou-Dalton transfer are not necessary required to obtain the incompatibility of the principles of aggregation and non-aggregation. Furthermore, as seen later, ratio aggregation is consistent with minimal non-aggregation and standard axioms. However, combining ratio aggregation with replication invariance still yields an impossibility result under the assumption of standard axioms. Since ratio aggregation implies the minimal aggregation principle proposed by Fleurbaey and Tungodden (2010), the following impossibility result is a corollary of their result (Fleurbaey and Tungodden 2010, Proposition 2).

**Proposition 2.** *There exists no social quasi-ordering that satisfies weak Pareto, Pigou-Dalton transfer, replication invariance, ratio aggregation, and minimal non-aggregation.*

Proof. Let $\bar{u} > \theta_r > \theta_p > \underline{u}$ and $N = \{1, \ldots, n\}$, where $\theta_p \geq \underline{u} - \delta + \alpha$ and $n - 1 > \lambda n$. Choose natural numbers $k, l \in \mathbb{N}$ such that $\underline{u} > \underline{u} - \delta + \frac{l\alpha}{k}$ and $\bar{u} > \bar{u} + \gamma - l\beta$. Then, consider the following profiles.

$$u_N = (\underline{u}, (n-1) * \bar{u});$$

$$u'_N = (\underline{u} - \delta, (n-1) * (\bar{u} + \gamma));$$

$$u_{k*N} = k * u_N = (k * \underline{u}, k * (n-1) * \bar{u});$$

$$u'_{k*N} = k * u'_N = (k * (\underline{u} - \delta), k(n-1) * (\bar{u} + \gamma));$$

$$w^0_{k*N} = (\underline{u} - \delta + \alpha, (k-1) * (\underline{u} - \delta), k(n-1) * (\bar{u} + \gamma - \beta));$$

$$w^1_{k*N} = (k * (\underline{u} - \delta + \tfrac{\alpha}{k}), k(n-1) * (\bar{u} + \gamma - \beta));$$

$$w^2_{k*N} = (k * (\underline{u} - \delta + \tfrac{2\alpha}{k}), k(n-1) * (\bar{u} + \gamma - 2\beta));$$

$$\vdots$$

$$w^l_{k*N} = (k * (\underline{u} - \delta + \tfrac{l\alpha}{k}), km * (\bar{u} + \gamma - l\beta)).$$

By ratio aggregation and $n - 1 \geq \lambda n$, $u'_N \succcurlyeq u_N$.

By replication invariance, $u'_{k*N} \succcurlyeq u_{k*N}$.

By minimal non-aggregation, $w^0_{k*N} \succcurlyeq u'_{k*N}$.

By Pigou-Dalton transfer, $w^1_{k*N} \succcurlyeq w^0_{k*N}$.

By the similar applications of minimal non-aggregation and Pigou-Dalton transfer, $w^l_{k*N} \succcurlyeq \cdots \succcurlyeq w^2_{k*N} \succcurlyeq w^1_{k*N}$.

By transitivity, $w^l_{k*N} \succcurlyeq u_{k*N}$.

However, weak Pareto implies $u_{k*N} > w^l_{k*N}$, which is a contradiction. ∎

Proposition 2 holds for any parameters of the axioms. In order to obtain this impossibility result, unlike Proposition 1, Pigou-Dalton transfer must be required. However, depending on the range of parameters, the following impossibility result can be obtained without requiring Pigou-Dalton transfer.

**Proposition 3.** *Suppose $\exists h, n \in \mathbb{N}, \ n > h[\lambda n] \ \& \ h\delta > \alpha$. Then, there exists no social quasi-ordering that satisfies weak Pareto, replication invariance, ratio aggregation, and minimal non-aggregation.*

Proof. Let $\bar{u} > \theta_r > \theta_p > \underline{u}$ and $N = \{1, \ldots, n\}$, where $\theta_p \geq \underline{u} + \alpha$. Choose $\bar{u}$ and natural numbers $n, h, k \in \mathbb{N}$ such that $n > h[\lambda n]$, $\bar{u} - k\beta > \theta_r$, $\underline{u} > \underline{u} + \alpha - h\delta$, and $\bar{u} > \bar{u} + \gamma - k\beta$. Then, consider the following profiles.

$u_{k*N} = (k * \underline{u}, k(n-1) * \bar{u})$;

$u^1_{k*N} = (\underline{u} + \alpha, (k-1) * \underline{u}, \ k(n-1) * (\bar{u} - \beta))$;

$u^2_{k*N} = (\underline{u} + \alpha, \underline{u} + \alpha, (k-2) * \underline{u}, k(n-1) * (\bar{u} - 2\beta))$;

⋮

$u^k_{k*N} = (k * (\underline{u} + \alpha), \ k(n-1) * (\bar{u} - k\beta))$;

$u_N = (\underline{u}, (n-1) * \bar{u})$;

$w_N = ((\underline{u} + \alpha), \ (n-1) * (\bar{u} - k\beta))$;

$$w_N^1 = (\underline{u} + \alpha - \delta,\ [\lambda n] * (\overline{u} + \gamma - k\beta), (n - 1 - [\lambda n]) * (\overline{u} - k\beta));$$

$$w_N^2 = (\underline{u} + \alpha - 2\delta,\ 2[\lambda n] * (\overline{u} + \gamma - k\beta), (n - 1 - 2[\lambda n]) * (\overline{u} - k\beta));$$

$$\vdots$$

$$w_N^h = (\underline{u} + \alpha - h\delta),\ h[\lambda n] * (\overline{u} + \gamma - k\beta), (n - 1 - h[\lambda n]) * (\overline{u} - k\beta)).$$

By minimal non-aggregation, $u_{k*N}^k \succcurlyeq \cdots \succcurlyeq u_{k*N}^2 \succcurlyeq u_{k*N}^1 \succcurlyeq u_{k*N}$.

By transitivity, $u_{k*N}^k \succcurlyeq u_{k*N}$.

Because $u_{k*N}^k = k * w_N$ and $u_{k*N} = k * u_N$, replication invariance implies $w_N \succcurlyeq u_N$.

By ratio aggregation, $w_N^h \succcurlyeq \cdots \succcurlyeq w_N^2 \succcurlyeq w_N^1 \succcurlyeq w_N$.

By transitivity, $w_N^h \succcurlyeq u_N$.

However, weak Pareto implies $u_N \succ w_N^h$, which is a contradiction. ∎

Note that the combination of ratio aggregation and replication invariance is independent of quantitative aggregation. As Propositions 2 and 3 simply show, the combination of ratio aggregation and replication invariance yields an impossibility result with the non-aggregation principle. In this sense, as Fleurbaey and Tungodden (2010) suggest with keen insight, the replication invariance axiom, which is considered as the most fundamental condition of population consistency in social choice theory with variable population, plays an important role in the impossibilities of the principles of aggregation and non-aggregation. Moreover, combining replication invariance with strong versions of non-aggregation would have catastrophic consequences for the aggregation problem. Indeed, strengthening the non-aggregation requirement in a way that does not take any thresholds into account, a combination of this strong non-aggregation and replication invariance implies that the following same-population leximin rule can survive under the standard axioms.

**The same-population leximin rule.**

A social quasi-ordering $\succcurlyeq^{SPL}$ is the *same-population leximin rule* iff $\forall u_N, v_N \in U$, $u_N \succcurlyeq^{SPL} v_N \leftrightarrow u_{[N]} = v_{[N]}$ or $u_{[1]} > v_{[1]}$ or $[\exists h \in \mathbb{N}$ with $h < n$, $(u_{[1]}, \ldots, u_{[h]}) = (v_{[1]}, \ldots, v_{[h]})$ and $u_{[h+1]} > v_{[h+1]}]$.

Note that this leximin rule says nothing about how to compare profiles with different population sizes. Indeed, for any profiles $u_N, v_M \in U$ with different population sizes, any social ranking on them is possible (including even an incomplete ranking). Proposition 4 shows that a combination of anonymity, strong Pareto, replication invariance, and strong non-aggregation is a necessary and sufficient condition for this leximin rule.

**Proposition 4.** *A social quasi-ordering satisfies anonymity, strong Pareto, replication invariance, and strong non-aggregation iff it is the same-population leximin rule.*

Proof. It is trivial that the same-population leximin rule satisfies the above axioms. Let's prove that a social quasi-ordering $\succcurlyeq$ that satisfies the above axioms must be the same-population leximin rule.

First, I would like to show that $\forall u_N, v_N \in U, \exists h \in \mathbb{N}, (u_{[1]}, \ldots, u_{[h]}) = (v_{[1]}, \ldots, v_{[h]})$ and $u_{[h+1]} > v_{[h+1]}$ imply $u_N \succ v_N$.

Obviously, whenever $u_{[h+1]} \geq v_{[n]}$, strong Pareto and anonymity imply $u_N \sim u_{[N]} \succ v_{[N]} \sim v_N$. By transitivity, $u_N \succ v_N$. Hence, suppose $u_{[h+1]} < v_{[n]}$. Consider $u^*, v^*, \alpha, \beta' \in \mathbb{R}$, and $k \in \mathbb{N}$ such that $v^* > v_{[n]} > u_{[h+1]} > v^* - k\beta' > u^* + \alpha > u^* > v_{[h+1]}$, where $\beta' \in (0, \beta)$ and $\beta$ is an acceptable sacrifice that is corresponding to $u^* + \alpha$. Then, define the following profiles:

$$w_{k*N}^0 = (k * (u_{[1]}, \ldots, u_{[h]}), k * u^*, k(n-h-1) * v^*);$$

$$w_{k*N}^1 = (k * (u_{[1]}, \ldots, u_{[h]}), u^* + \alpha, (k-1) * u^*, k(n-h-1) * (v^* - \beta'));$$

$$w_{k*N}^2 = (k * (u_{[1]}, \ldots, u_{[h]}), 2 * (u^* + \alpha), (k-2) * u^*, k(n-h-1) * (v^* - 2\beta'));$$

$$\vdots$$

$$w_{k*N}^l = (k * (u_{[1]}, \ldots, u_{[h]}), k * (u^* + \alpha), k(n-h-1) * (v^* - k\beta')).$$

By strong non-aggregation, $w_{k*N}^l \succcurlyeq \cdots \succcurlyeq w_{k*N}^1 \succcurlyeq w_{k*N}^0$.

By strong Pareto and anonymity, $k * u_N \sim k * u_{[N]} \succ w_{k*N}^l$ and $w_{k*N}^0 \succ k * v_{[N]} \sim k * v_N$.

By transitivity, $k * u_N \succ k * v_N$.

By replication invariance, $u_N \succ v_N$.

Finally, by the similar logic of the above argument, it is easily established that $\forall u_N, v_N \in U, \ u_{[1]} > v_{[1]} \rightarrow u_N \succ v_N$.

Thus, $\succcurlyeq$ must be the same-population leximin rule. ∎

This result is interesting because it provides a new characterization of the leximin rule in terms of the non-aggregation principle. In fact, traditional characterization results of the leximin rule are divided into three categories: the first category is related to the rank dictatorship, which is obtained by combining ordinal full interpersonal comparability of well-being with standard axioms (d'Aspremont and Gevers 1977); the second category is related to combining standard axioms with extreme progressive transfer principles such as the Hammond equity axiom (Hammond 1976; d'Aspremont and Gevers 1977); the third category is related to the joint characterization in a reduced form (Dechamps and Gevers 1978; Sakamoto 2024), in which combining standard axioms with scale invariance for positive affine transformations yields the weak utilitarian and leximin rules. Proposition 4 provides the fourth category of a new characterization of the leximin rule in the sense that the leximin property can be obtained by combining a stronger version of the non-aggregation principle with replication invariance. Thus, under the standard assumptions in social choice theory (anonymity, strong Pareto and social quasi-ordering), the combination of strong non-aggregation and replication invariance goes far beyond the role of the original non-aggregation principle. The combination of these axioms implies that, given a population size, a slight improvement in well-being of just one individual has priority over a large sacrifice in well-being of a large number of well-off individuals.

      As Proposition 2 shows, when one tries to apply the non-aggregation principle without any consideration of thresholds, replication invariance excludes compatibility with any form of the aggregation principle. What then happens in the case of the strong non-aggregation principle[9] that takes the threshold levels into account? Stepwise social

---

[9] The strong non-aggregation principle that considers the threshold constraints is defined as follows.

**Strong Non-Aggregation with Threshold Constraints.**

welfare orderings, a class of practical and acceptable social welfare orderings that Sakamoto and Mori (2021) proposed, clarifies this question. In general, a class of social welfare ordering satisfying replication invariance is represented by a value function $W(n, V(u_N))$ where $n$ is the population size and $V(u_N)$ is a representative well-being of the profile $u_N$. The nested value function $V(u_N)$ can be defined as the aggregated form of a stepwise function that transforms the profile $u_N$ with the discrete population into a continuous closed interval [0, 1]. However, a marginal contribution that an individual with well-being below a threshold can make to $V(u_N)$ must be proportional to the reciprocal of the population size, $1/n$. This means that the marginal contribution of this single person's well-being improvement converges to zero as the population size $n$ increases. Therefore, as long as replication invariance is requested, even if one tries to apply the non-aggregation principle only to individuals below the threshold, it will be necessary to give them absolute priority to make their marginal contributions overwhelm well-being of the remaining population $(n-1)/n$. As a result, it is intuitively understood that this is incompatible with any type of the aggregation principle. For example, a *sufficientarian social welfare ordering* satisfies the strong non-aggregation principle with threshold constraints and all standard axioms except continuity, but it violates the aggregation principle.

    Let us consider the compatibility problem of the principles of aggregation and non-aggregation. As Proposition 1 shows, quantitative aggregation and minimal non-aggregation are inconsistent even without imposing replication invariance. Fleurbaey and

---

$\exists \theta_r > \theta_p > 0$, $\forall \alpha > 0$, $\exists \beta > 0$ with $\alpha > \beta, \forall u_N, v_N \in U$, $\forall i \in N$, $\forall M \subseteq N \setminus \{i\}$, if $[\forall j \in M, u_j - \beta = v_j \geq \theta_r > \theta_p \geq v_i = u_i + \alpha]$ & $[\forall h \in N \setminus (M \cup \{i\}), v_h = u_h]$, then $v_N \succcurlyeq u_N$.

Tungodden (2010, Proposition 1), however, find a possibility result if replication invariance is dropped and the aggregation principle is restricted to the domain with some fixed population. A social ordering that they proposed is the *geometric Gini ordering*, but in this paper, I will examine a generalization of it, the *rank-discounted generalized utilitarian rule*.[10]

**Rank-Discounted Generalized Utilitarianism.**

$\forall u_N, v_N \in U, \ u_N \succcurlyeq v_N \leftrightarrow \sum_{[i]\in[N]} \rho^{-([i]-1)} g(u_{[i]}) \geq \sum_{[i]\in[N]} \rho^{-([i]-1)} g(v_{[i]})$.

Rank-discounted generalized utilitarianism is a variant of rank-weighted generalized utilitarianism. Rank-dependent weights are given as a power of the discount factor $\rho$. The worst-off individual is given a weight of 1, the second worst-off individual is given a weight $\rho^{-1}$, and similarly, the best-off individual is given a weight of $\rho^{-(n-1)}$. As is clear from the definition, if the discount factor is greater than 1, this social ordering tends to ignore well-being of individuals with higher ranks for sufficiently large population sizes. Therefore, it is only for relatively small populations that the aggregation principle can be satisfied, and it is not expected to satisfy the aggregation principle for sufficiently large populations. Needless to say, this social ordering violates quantitative aggregation from Proposition 1 because it satisfies strong Pareto, Pigou-Dalton transfer, and minimal non-aggregation (under appropriate settings of discount factors and concave functions). Furthermore, as the following proposition shows, this social ordering cannot satisfy the ratio aggregation principle.

---

[10] A class of rank-discounted generalized utilitarian orderings has been studied in terms of axiomatic characterizations and theoretical properties in the contexts of social choice with both variable and infinite populations. See, for example, Zuber and Asheim (2012), Asheim and Zuber (2014; 2022). Especially, Asheim and Zuber (2022, Propositions 4 and 5) analyze theoretical relationships between the standard axioms (including replication invariance) and minimal versions of the aggregation and non-aggregation principles in a limited class of social rankings.

**Proposition 5.** *For all $\rho > 1$, a rank-discounted generalized utilitarian rule satisfies minimal non-aggregation if and only if $g(\theta_p) - g(\theta_p - \alpha) \geq (\rho - 1)^{-1}\rho[g(\theta_r + \beta) - g(\theta_r)]$. Moreover, if $\rho > 1$, then a rank-discounted generalized utilitarian rule violates ratio aggregation.*

Proof. Consider a situation in which well-being of an individual with the lowest rank increases by $\alpha$ and remains below the threshold $\theta_p$, while well-being of all the remaining individuals decreases by $\beta$ and remains at a level above the threshold $\theta_r$. Due to the concavity of the function $g$, the minimum increment in social welfare in rank-discounted generalized utilitarianism by the increase in well-being of the individual with the lowest rank is given by $g(\theta_p) - g(\theta_p - \alpha)$. The maximum decrease in social welfare in rank-discounted generalized utilitarianism by the decrease in well-being of all remaining individuals is given by $g(\theta_r + \beta) - g(\theta_r)$. The minimal aggregation principle prevents this change in the well-being profile from decreasing social welfare, so the following must hold for all $n \in \mathbb{N}$.

$$g(\theta_p) - g(\theta_p - \alpha) \geq \sum_{i=1}^{n} \rho^{-(i-1)}[g(\theta_r + \beta) - g(\theta_r)],$$
$$\leftrightarrow g(\theta_p) - g(\theta_p - \alpha) \geq \lim_{n \to \infty} \sum_{i=1}^{n} \rho^{-(i-1)}[g(\theta_r + \beta) - g(\theta_r)],$$
$$\leftrightarrow g(\theta_p) - g(\theta_p - \alpha) \geq (\rho - 1)^{-1}\rho[g(\theta_r + \beta) - g(\theta_r)].$$

The left-hand side corresponds to the minimum increase in social welfare, and the right-hand side corresponds to the maximum increase in social welfare, so if the above equation holds, then minimal non-aggregation is always satisfied for any profile.

Next, let $\rho > 1$ and $\lambda \in (0,1)$. I will show that rank-discounted generalized utilitarianism does not satisfy the ratio aggregation principle. If rank-discounted generalized utilitarianism could satisfy ratio aggregation, the following must hold true for all profiles $u_N$.

$$g(u_{[1]}) - g(u_{[1]} - \delta) \leq \sum_{i=n-[\lambda n]+1}^{n} \rho^{-i} [g(u_{[i]} + \gamma) - g(u_{[i]})]$$

$$\leq \sum_{i=n-[\lambda n]+1}^{n} \rho^{-i} [g(u_{[n-[\lambda n]+1]} + \gamma) - g(u_{[n-[\lambda n]+1]})],$$

$$\leftrightarrow g(u_{[1]}) - g(u_{[1]} - \delta) \leq \frac{\rho^{-n+[\lambda n]-1} - \rho^{-n-1}}{\rho - 1} [g(u_{[n-[\lambda n]+1]} + \gamma) - g(u_{[n-[\lambda n]+1]})].$$

The left-hand side of the above equation always takes a positive value, and the difference term $g(u_{[n-[\lambda n]+1]} + \gamma) - g(u_{[n-[\lambda n]+1]})$ on the right-hand side also takes a positive value. However, since the limit of the coefficient $\rho^{-n+[\lambda n]-1} - \rho^{-n-1}$ on the right-hand side is zero, the above inequality no longer holds true for a sufficiently large natural number $n^*$. Thus, rank-discounted generalized utilitarianism violates ratio aggregation. ∎

The necessary and sufficient condition for satisfying the minimal non-aggregation principle in Proposition 5 is extremely intractable for the compatibility of the aggregation principle. Since the function $g$ is monotonically increasing, continuous, and concave, it follows that even if the left-hand side is sufficiently large, the function $g$ must be adjusted so that it is larger than the right-hand side, and the coefficient $(\rho - 1)^{-1} \rho$ must also be sufficiently small. However, a very small $(\rho - 1)^{-1} \rho$ means that $\rho$ must be sufficiently large, which means that the rank-dependent weights of individuals with higher ranks must decrease geometrically. Thus, as population size increases, well-being of individuals with higher ranks will inevitably be ignored, and large improvements in well-being of a certain percentage of the population will be overwhelmed by a tiny sacrifice of individuals with the lowest rank. In fact, the rank-discounted generalized utilitarian ordering cannot satisfy the ratio aggregation principle. This seems to be a major flaw in the rank-discounting approach. One of the major problems with the maximin and leximin rules is the so-called leveling-down objection, and rank-discounted generalized utilitarianism has a similar problem. For example, let us consider a well-being profile comparison problem with a population of 100 million and a discount factor of 1.01. In this case, the well-being level of the 99% of the population after rank 1 million would be almost worthless, and an irrational judgment would be recommended that calls for a large sacrifice of the 99% of the population for a small gain of one person with the lowest rank.[11]

---

[11] If the discount factor is 1.01, well-being of individuals with higher ranks, even after rank 1,000, become almost worthless. For example, in rank-discounted generalized utilitarianism with a simple concave function such as $\forall u \in \mathbb{R}, g(u) = \sqrt{u}$, a profile $u_N = (10^6 * 100)$ is socially better than $v_N = (90, (10^3 - 1) * 100, (10^6 - 10^3) * 300)$. That is, the small benefit of the worst-off person overwhelms the great benefits of the top 99% persons. Furthermore, note that rank-discounted generalized utilitarianism also raises problems similar to the *reverse repugnant conclusion* in population ethics. For example, in the above rank-discounted generalized utilitarianism, a profile $u_N = (10^3 * 100)$ is socially better than $v_N = (10^6 * 99)$. That is, small populations with slightly higher well-being would be socially better than enormous populations with slightly lower well-being

Under the assumption of standard axioms such as anonymity, strong Pareto, and Pigou-Dalton transfer, is there any reasonable social ordering that would satisfy both ratio aggregation and non-aggregation? The answer is yes. In fact, let us define the following practical social ordering that satisfies ratio aggregation while preserving the spirit of the non-aggregation principle -- it is unacceptable to impose great sacrifice on one individual for the sake of a small gain for many wealthier individuals -- holds.

**Proposition 6.** *Suppose $\alpha = \gamma$ and $\beta = \delta$ in the definitions of the principles of ratio aggregation and minimal non-aggregation. There exists a social ordering that satisfies anonymity, strong Pareto, Pigou-Dalton transfer, ratio aggregation, and minimal non-aggregation.*[12]

Proof. For example, consider the following population-dependent value function $V^n$ which obviously satisfies the above axioms.

$\exists \theta_p > 0$, $\forall n \in \mathbb{N}$, $\exists \lambda^n \in (0,1)$, $\forall u_N \in U$ with $|N| = n$, $V^n(u_N) = \lambda^n \sum_{i: u_i < \theta_p}[u_i - \theta_p] + (1 - \lambda^n)\frac{1}{n}\sum_{i \in N} u_i$,

---

but sufficiently good lives.

[12] In fact, the social ordering in Proposition 6 satisfies a stronger requirement of the non-aggregation principle that is defined as follows: $\exists \theta_p > 0$, $\exists \alpha > \beta > 0$, $\forall u_N, v_N \in U, \forall i \in N$, $\forall M \subseteq N \setminus \{i\}$, if $\left[\theta_p \geq v_i \geq u_i + \alpha\right]$ & $\left[\forall j \in M, v_j \geq u_j - \beta \geq \theta_p\right]$ & $\left[\forall h \in N \setminus (M \cup \{i\}), v_h = u_h\right]$, then $v_N \succcurlyeq u_N$.

where $\lambda^n$ is defined so as to satisfy both the principles of ratio aggregation and minimal non-aggregation, that is, $-\lambda^n \delta + (1-\lambda^n)\frac{1}{n}(\lceil \lambda n \rceil \gamma - \delta) \geq 0$ and $\lambda^n \alpha + (1-\lambda^n)\frac{1}{n}(\alpha - (n-1)\beta) \geq 0$ for all $n \in \mathbb{N}$.

Note that the above two inequalities are equivalent to the following one.

$$\frac{(n-1)\beta - \alpha}{(n-1)(\alpha + \beta)} \leq \lambda^n \leq \frac{\lceil \lambda n \rceil \gamma - \delta}{(n-1)\delta + \lceil \lambda n \rceil \gamma}$$

If $\alpha = \gamma$ and $\beta = \delta$, the above inequality holds true. Thus, there exists a class of *population-dependent rank-weighted generalized utilitatian rules* $\{g^n\}_{n \in \mathbb{N}}$ which satisfy the above axioms. ∎

This social ordering is defined as a linear combination of the sufficientarian sum (a total value that adds up well-being of individuals below a certain threshold) and simple average. By definition, the value function representing this social ordering can be interpreted as a generalized utilitarianism that changes its functional form with population size. Indeed, if individual $i$'s well-being is below the threshold $\theta_p$, then the individual's well-being is valued at $\lambda^n(u_i - \theta_p) + (1 - \lambda^n)u_i/n$ and if it is above the threshold $\theta_p$, then the individual's well-being is valued at $(1 - \lambda^n)u_i/n$. Therefore, by defining the function $g^n(u_i)$ as follows, this social ordering can be viewed as a generalized utilitarianism that changes its functional form with population size.

$$g^n(u_i) = \begin{cases} \lambda^n(u_i - \theta_p) + (1-\lambda^n)\frac{1}{n}u_i, & u_i < \theta_p \\ (1-\lambda^n)\frac{1}{n}u_i, & u_i \geq \theta_p \end{cases}$$

In this functional form, all that the social planner has to decide on will be the values of parameter $\lambda^n$, which indicates the degree to which the non-aggregation principle is respected, and the threshold of poverty $\theta_p$. As will be remarked in my conclusion, it is easy to generalize this social ordering to various forms, and further axiomatic research will be needed.[13]

---

[13] As is well-known, generalized utilitarianism can be characterized by anonymity, strong Pareto, continuity, and separability. However, it will be necessary to examine a subclass of the generalized utilitarian orderings presented in Proposition 6 in terms of both the principles of aggregation and non-aggregation. While there are various social orderings that satisfy many types of the non-aggregation principles, the author does not currently have a firm answer to the question of which class to be chosen.

# 4. Concluding Remarks

This paper revisits impossibility results on the tyrannies of aggregation and non-aggregation. I showed that quantitative aggregation and minimal non-aggregation are incompatible, and that the combination of strong non-aggregation and replication invariance leads to the leximin rule under the assumption of standard axioms. On the other hand, I also showed that ratio aggregation is compatible with minimal non-aggregation by dropping replication invariance. It is practically significant that there is a class of acceptable social orderings that satisfies the standard axioms and avoids the tyrannies of aggregation and non-aggregation. This social ordering would allow us to properly treat great benefits of minorities in the pharmacoeconomic evaluation problem. It is also expected to be applicable to the issue of economic evaluation of reasonable accommodation for persons with disabilities. Needless to say, the results of this study are only on the first step in the development of practical evaluation methods, and further research is needed.

First, it is easy to modify the social ordering presented in Proposition 6 to take into account not only a specific threshold but also multiple thresholds. For example, consider the following social ordering.

$\exists (\theta_1, \ldots, \theta_{k-1})$ with $\theta_1 < \cdots < \theta_{k-1}$, $\forall n \in \mathbb{N}$, $\exists \lambda_j^n \in (0,1)$ with $\lambda_1^n > \cdots > \lambda_k^n$ and $\sum_{j=1}^{k} \lambda_j^n = 1$, $\forall u_N \in U$ with $|N| = n$, $V^n(u_N) = \sum_{j=1}^{k-1} \sum_{i:\, u_i < \theta_j} \lambda_j^n [u_i - \theta_j] + \lambda_k^n \frac{1}{n} \sum_{i \in N} u_i$ where $\theta_0 = -\infty$.

Note that although this social ordering is designed to satisfy the non-aggregation principle across multiple threshold intervals, it is still a variant of generalized utilitarianism.

Second, although the social ordering presented in this study does not take relative inequality into account, it is possible to have a consideration on relative inequality by transforming the second term into the rank-weighted utilitarianism. The following social ordering is one of examples considering relative inequality.

$\exists \theta_p > 0, \ \forall n \in \mathbb{N}, \ \exists \lambda^n \in (0,1), \ \forall u_N \in U$ with $|N| = n, \ V^n(u_N) = \lambda^n \sum_{i:\ u_i < \theta_p}[u_i - \theta_p] + (1 - \lambda^n) \sum_{[i] \in [N]} w_{[i]} u_{[i]}$.

Third, it should not be acceptable to impose a great sacrifice on one individual for the greater happiness of all others. However, the social ordering in Proposition 6 may promote this morbid judgment. For example, suppose that the parameter $\lambda$ of this social ordering is 0.2 and its threshold $\theta_p$ is 0 in a nation called Omelas with a population size of 100 million. In this case, a profile $u_N = (-100, (10^9 - 1) * 200)$ is socially better than $v_N = (0, (10^9 - 1) * 100)$, which means that great gains of many individuals have priority over a great sacrifice of one individual. To avoid such a problem, an increasing, continuous, and concave function $g$ with its upper bound may be useful. Note, of course, that using such a function will yield a new firestorm in which great gains for individuals who are better-off above a certain level will be almost worthless.[14]

---

[14] Several counterarguments can be made regarding this issue. First, human cognitive senses have limits in their discernible threshold and acceptable range, so neither senses nor emotions respond

$\exists \theta_p > 0$, $\forall n \in \mathbb{N}$, $\exists \lambda^n \in (0,1)$, $\forall u_N \in U$ with $|N| = n$, $V^n(u_N) = \lambda^n \sum_{i: u_i < \theta_p}[u_i - \theta_p] + (1-\lambda^n)\frac{1}{n}\sum_{i \in N} g(u_i)$,

where a function $g$ is increasing, continuous, and concave and has its upper bound.

Forth, a social planner may think that the first term of the social ordering in Proposition 6 should be modified so as to take a consideration of the relatively poor's well-being below the threshold. Then, the following one is an example of such a social ordering.

$\exists$ concave function $g$, $\exists \theta_p > 0$, $\forall n \in \mathbb{N}$, $\exists \lambda^n \in (0,1)$, $\forall u_N \in U$ with $|N| = n V^n(u_N) = \lambda^n \sum_{i: u_i < \theta_p}[g(u_i) - g(\theta_p)] + (1-\lambda^n)\frac{1}{n}\sum_{i \in N} u_i$.

Finally, in order to obtain a practicable and acceptable evaluation method of social welfare, estimating the functional form and parameters of these social orderings is necessary. However, such work will require experimental methods in a fair information environment. Such experimental and empirical studies will be needed in the future in order to construct a practicable and acceptable social evaluation.

---

sharply to changes in external conditions. Therefore, there are probably upper and lower limits to human levels of well-being and feelings of happiness. In this case, since well-being level is bounded, extreme numerical examples like Omelas may not be possible. Second, when evaluating actual public policy, it is rare that a great sacrifice of one individual results in significant improvements for the majority. Of course, the question of how much public fund should be spent on infrastructure development in remote areas may be similar to this problem. However, because the residents in remote areas can generally move to the center of town, it doesn't make the problem any more serious. Third, Omelas's numerical example may be avoided if the parameters are set appropriately.